\documentclass[a4paper,12pt]{article}

\usepackage{ifpdf}
\newif\ifpdf
\ifx\pdfoutput\undefined
  \pdffalse
\else
  \pdfoutput=1
  \pdftrue
\fi

\RequirePackage{xspace} %
\RequirePackage{subfigure} %
\RequirePackage[centertags]{amsmath} %
\RequirePackage{amssymb}
\RequirePackage{wrapfig} %
\RequirePackage{calc} %
\RequirePackage{ifthen}
\RequirePackage{tabularx} %
\RequirePackage{flafter} %
\RequirePackage{fancyhdr} %

\ifpdf
  \RequirePackage[pdftex]{color}%
  \RequirePackage{colortbl}%
  \RequirePackage{array}%
  \RequirePackage[pdftex]{graphicx}
  \RequirePackage[ pdftex, plainpages = false, pdfpagelabels,
                 pdfpagelayout = useoutlines,
                 bookmarks,
                 breaklinks = true,
                 linktocpage,
                 pagebackref,                      
                 colorlinks = true,
                 linkcolor = blue,
                 urlcolor  = blue,
                 citecolor = blue,
                 anchorcolor = blue,
                 hyperindex = true,
                 hyperfigures
                 ]{hyperref}

\else
  \RequirePackage{color}
  \RequirePackage{colortbl}
   \RequirePackage{array}
  \RequirePackage[dvips]{graphicx}
  \RequirePackage{hyperref}
  \usepackage{rotating}
\fi

\usepackage{makeidx} %
\usepackage{setspace} %
\usepackage{rotating} %
\usepackage{ecltree}
\usepackage{epic}
\usepackage{supertabular}  %
\usepackage{color}
\usepackage{exscale}
\usepackage{fontenc}
\usepackage{ifthen}
\usepackage{latexsym}
\usepackage{makeidx}
\usepackage{syntonly}
\usepackage{inputenc}
\usepackage{graphicx}
\usepackage{setspace}
\usepackage{caption2}
\usepackage[english]{babel}
\usepackage[square, comma,numbers,sort&compress]{natbib}
\usepackage{hypernat}
\usepackage{boxedminipage}
\usepackage{framed}
\usepackage{longtable}
\usepackage[all]{hypcap}

\newcommand{\note}[1]{\marginpar[left]{\singlespace \tiny #1}}
\renewcommand{\sectionmark}[1]%
      {\markright{\thesection\ #1}} 

\renewcommand{\note}[1]{}

\newcommand{\sR}     {\dot{\gamma}}         %
\newcommand{\sS}     {\tau}           
\newcommand{\arccosh}     {{\rm arccosh}}

\doublespace 

\title
{ %
\vspace*{3.0cm} \LARGE{\bf The Flow of Power-Law Fluids in Axisymmetric Corrugated Tubes} \vspace*{4.0cm} \\
}

\author{Taha Sochi\footnote{University College London, Department of Physics \& Astronomy, Gower Street, London, WC1E 6BT.
Email: t.sochi@ucl.ac.uk.} \vspace*{5.0cm}}

\begin{document}

\maketitle %
\pagenumbering{arabic}

\newpage
\phantomsection \addcontentsline{toc}{section}{Contents} %
\tableofcontents

\newpage
\phantomsection \addcontentsline{toc}{section}{List of Figures} %
\listoffigures

\newpage
\phantomsection \addcontentsline{toc}{section}{Abstract} \noindent
{\noindent \LARGE \bf Abstract} \vspace{0.5cm}\\
\noindent %

In this article we present an analytical method for deriving the relationship between the pressure
drop and flow rate in laminar flow regimes, and apply it to the flow of power-law fluids through
axially-symmetric corrugated tubes. The method, which is general with regards to fluid and tube
shape within certain restrictions, can also be used as a foundation for numerical integration where
analytical expressions are hard to obtain due to mathematical or practical complexities. Five
converging-diverging geometries are used as examples to illustrate the application of this method.

\pagestyle{headings} %
\addtolength{\headheight}{+1.6pt}
\lhead[{Chapter \thechapter \thepage}]%
      {{\bfseries\rightmark}}
\rhead[{\bfseries\leftmark}]
     {{\bfseries\thepage}}
\headsep = 1.0cm

\newpage
\section{Introduction} \label{Introduction}

\begin{figure}[!h]
\centering{}
\includegraphics
[scale=0.6] {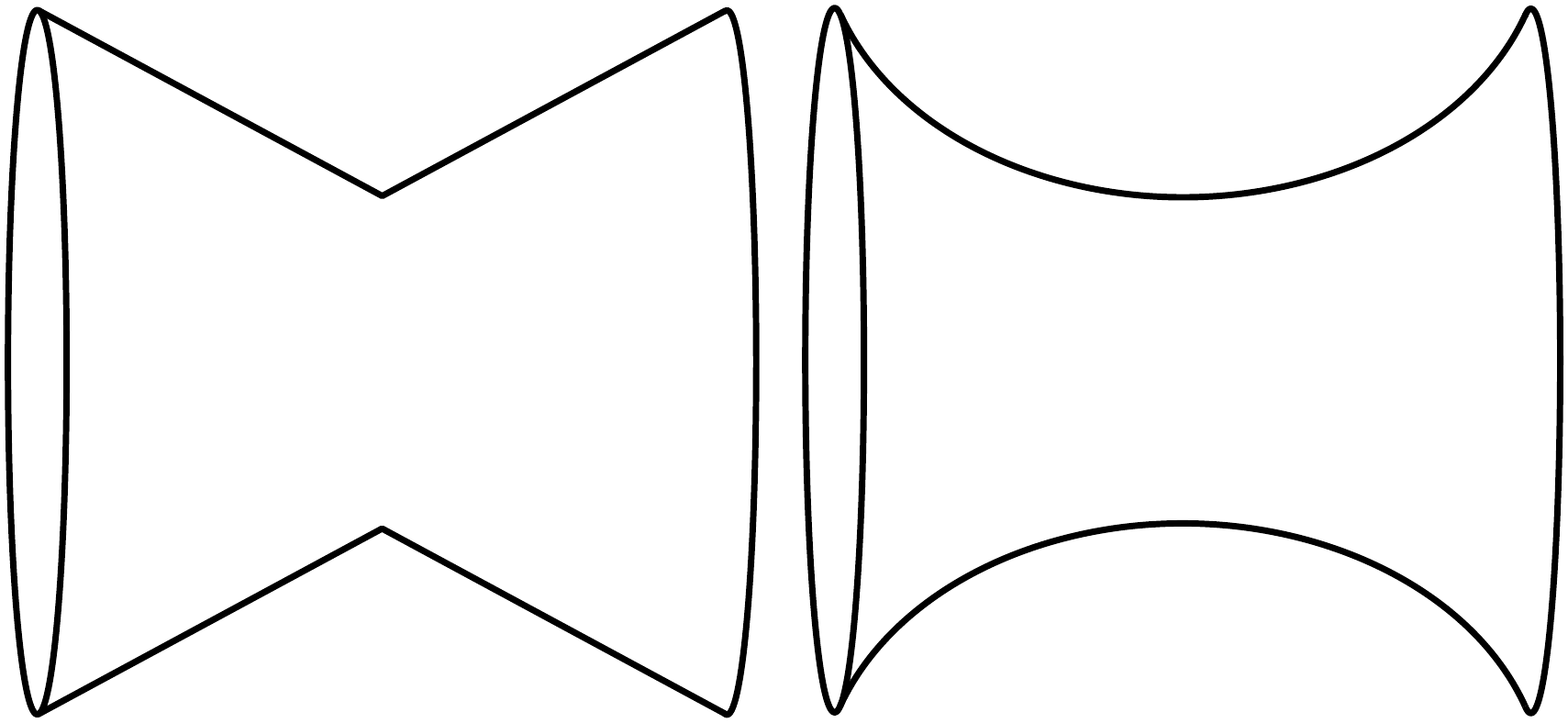} \caption{Profiles of converging-diverging axisymmetric capillaries.}
\label{Tubes}
\end{figure}

Modeling the flow in corrugated tubes is required for a number of technological and industrial
applications. Moreover, it is a necessary condition for correct description of several phenomena
such as yield-stress, viscoelasticity and the flow of Newtonian and non-Newtonian fluids through
porous media \cite{PilitsisB1989, TalwarK1992, SochiVE2009, SochiYield2010}. In the literature of
fluid mechanics there are numerous studies on the flow through tubes or channels of corrugated
nature. Most of these studies use numerical techniques such as spectral and finite difference
methods \cite{ThienGB1985, ThienK1987, LahbabiC1986, BurdetteCAB1989, PilitsisB1989, PilitsisB1992,
JamesTKBP1990, MasulehP2004}. Some others adopt analytical approaches based on simplified
assumptions and normally deal with very special cases \cite{Oka1973, WilliamsJ1980}. The current
paper presents a mathematical method for deriving analytical relations between the volumetric flow
rate and pressure drop in corrugated circular capillaries of variable cross section, such as those
illustrated schematically in Figure \ref{Tubes}. It also presents five examples in which this
method is used to derive flow equations for non-Newtonian fluids of power-law rheology. The method,
however, is more general and can be used for other tube shapes and other rheologies. The following
derivations assume a laminar flow of a purely-viscous incompressible fluid where the tube
corrugation is smooth and relatively small to avoid complex flow phenomena which are not accounted
for in the underlying assumptions of this method.

\section{$P-Q$ Relation for Power-Law Fluids}

\begin{figure} [!h]
\centering{}
\includegraphics
[scale=1] {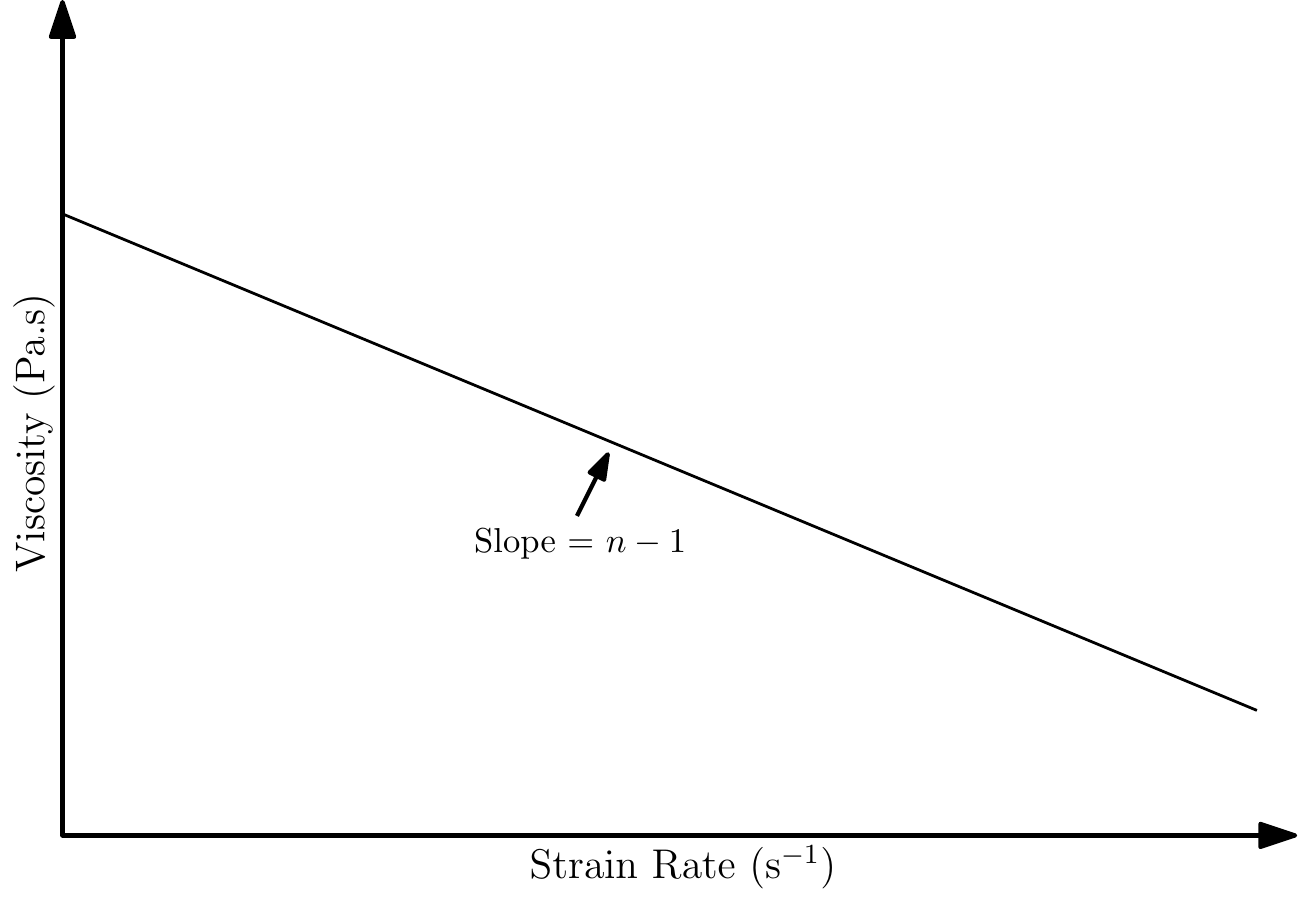} \caption{The bulk rheology of power-law fluids on logarithmic scales for
finite strain rates ($\sR>0$).} \label{PowLawRhe}
\end{figure}

The power-law, or Ostwald-de Waele, is a widely-used fluid model to describe the rheology of shear
thinning fluids. It is one of the simplest time-independent non-Newtonian models as it contains two
parameters only. The model is given by the relation \cite{BirdbookAH1987, CarreaubookKC1997}
\begin{equation} \label{strStrRel}
\mu = \frac{\sS}{\sR} = C\dot{\gamma}^{n-1}
\end{equation}
where $\mu$ is the fluid viscosity, $\sS$ is the stress, $\sR$ is the strain rate, $C$ is the
consistency factor, and $n$ is the flow behavior index. In Figure \ref{PowLawRhe} the bulk rheology
of this model for shear-thinning case is presented generically on log-log scales. Although the
power-law is used to model shear-thinning fluids, it can also be used for modeling shear-thickening
by making $n$ greater than unity. The major weakness of power-law model is the absence of plateaux
at low and high strain rates. Consequently, it fails to produce sensible results in these flow
regimes.

The volumetric flow rate, $Q$, of a power law fluid through a circular capillary of constant radius
$r$ with length $x$ across which a pressure drop $P$ is applied is given by

\begin{equation}
Q=\frac{\pi nr^{4}P^{1/n}}{2xC^{1/n}(3n+1)}\left(\frac{2x}{r}\right)^{1-1/n}
\end{equation}
A derivation of this relation, which can be obtained from Equation \ref{strStrRel}, can be found in
\cite{SochiB2008, Sochithesis2007}. On solving this equation for $P$ the following relation is
obtained
\begin{equation}
P=\frac{2CQ^{n}(3n+1)^{n}x}{\pi^{n}n^{n}r^{3n+1}}
\end{equation}

For an infinitesimal length, $\delta x$, of a capillary, the infinitesimal pressure drop for a
given flow rate $Q$ is

\begin{equation}
\delta P=\frac{2CQ^{n}(3n+1)^{n}\delta x}{\pi^{n}n^{n}r^{3n+1}}
\end{equation}

For an incompressible fluid, the volumetric flow rate across an arbitrary cross section of the
capillary is constant. Therefore, the total pressure drop across a capillary of length $L$ with
circular cross section of varying radius, $r(x)$, is given by
\begin{equation} \label{PQEq}
P=\frac{2CQ^{n}(3n+1)^{n}}{\pi^{n}n^{n}}\int_{0}^{L}\frac{dx}{r^{3n+1}}
\end{equation}

In the following sections, Equation \ref{PQEq} will be used to derive analytical expressions for
the relation between pressure drop and volumetric flow rate for five converging-diverging
geometries of axially-symmetric pipes. The method can be equally applied to other geometries as
long as the tube radius can be expressed analytically as a function of the tube axial distance.

\subsection{Conic Tube} \label{}

\begin{figure}[!h]
\centering{}
\includegraphics
[scale=0.8] {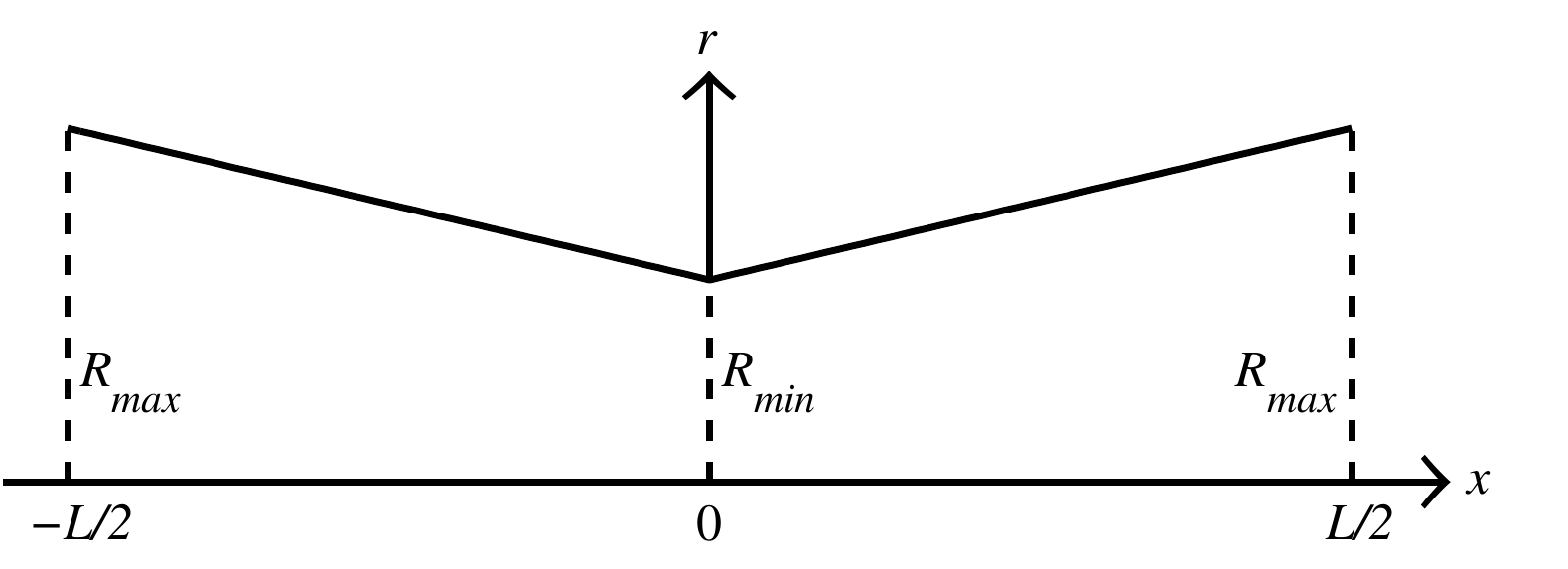} \caption{Schematic representation of the radius of a conically shaped
converging-diverging capillary as a function of the distance along the tube axis.} \label{Conic}
\end{figure}

For a corrugated tube of conic shape, depicted in Figure \ref{Conic}, the radius $r$ as a function
of the axial coordinate $x$ in the designated frame is given by

\begin{equation}
r(x)=a+b|x| \hspace{1cm} -L/2\le x \le L/2 \hspace{1cm} a,b>0
\end{equation}
where
\begin{equation}
a=R_{min} \hspace{1cm} {\rm and} \hspace{1cm} b=\frac{2(R_{max}-R_{min})}{L}
\end{equation}

Hence, Equation \ref{PQEq} becomes

\begin{eqnarray}
  P&=&\frac{2CQ^{n}(3n+1)^{n}}{\pi^{n}n^{n}}\int_{-L/2}^{L/2}\frac{dx}{(a+b|x|)^{3n+1}} \\
  &=&\frac{2CQ^{n}(3n+1)^{n}}{\pi^{n}n^{n}}\left(\left[\frac{1}{3bn(a-bx)^{3n}}\right]_{-L/2}^{0}+\left[-\frac{1}{3bn(a+bx)^{3n}}\right]_{0}^{L/2}\right) \\
  &=&\frac{4CQ^{n}(3n+1)^{n}}{3\pi^{n}n^{n+1}b}\left[\frac{1}{a^{3n}}-\frac{1}{(a+bL/2)^{3n}}\right]
\end{eqnarray}
that is

\begin{equation}\label{fConic}
P=\frac{2LCQ^{n}(3n+1)^{n}}{3\pi^{n}n^{n+1}(R_{max}-R_{min})}\left[\frac{1}{R_{min}^{3n}}-\frac{1}{R_{max}^{3n}}\right]
\end{equation}

\subsection{Parabolic Tube} \label{}

\begin{figure}[!h]
\centering{}
\includegraphics
[scale=0.8] {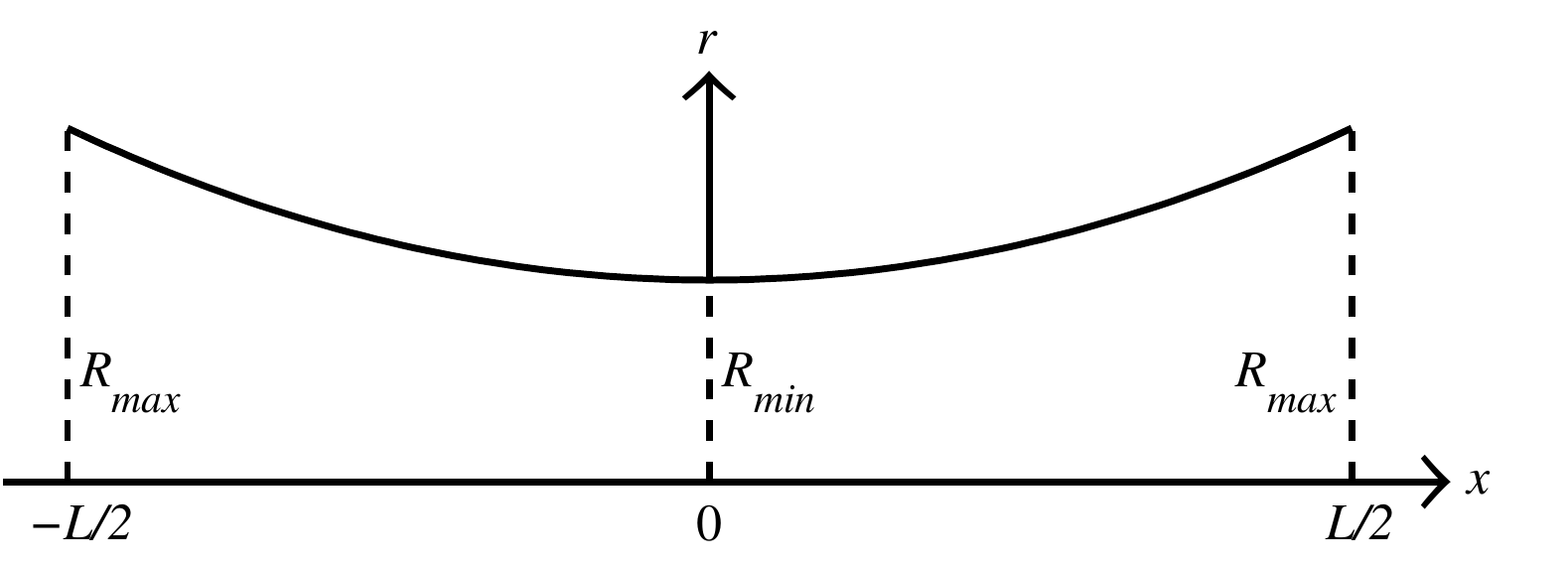} \caption{Schematic representation of the radius of a converging-diverging
capillary with a parabolic profile as a function of the distance along the tube axis.}
\label{Parabola}
\end{figure}

For a tube of parabolic profile, depicted in Figure \ref{Parabola}, the radius is given by

\begin{equation}
r(x)=a+bx^{2} \hspace{1cm} -L/2 \le x \le L/2 \hspace{1cm} a,b>0
\end{equation}
where
\begin{equation}
a=R_{min} \hspace{1cm} {\rm and} \hspace{1cm} b=\left(\frac{2}{L}\right)^{2}(R_{max}-R_{min})
\end{equation}

Therefore, Equation \ref{PQEq} becomes

\begin{eqnarray}
  P&=&\frac{2CQ^{n}(3n+1)^{n}}{\pi^{n}n^{n}}\int_{-L/2}^{L/2}\frac{dx}{\left(a+bx^{2}\right)^{3n+1}} \\
  &=&\frac{2CQ^{n}(3n+1)^{n}}{\pi^{n}n^{n}a}\left[x\left(\frac{\frac{bx^{2}}{a}+1}{a+bx^{2}}\right)^{3n}{}_{2}F_{1}\left(\frac{1}{2},3n+1;\frac{3}{2};-\frac{bx^{2}}{a}\right)\right]_{-L/2}^{L/2}
\end{eqnarray}
where $_{2}F_{1}$ is the hypergeometric function. Therefore

\begin{equation}\label{fParabolic}
P=\frac{2LCQ^{n}(3n+1)^{n}}{n^{n}\pi^{n}R_{min}^{3n+1}}\,_{2}F_{1}\left(\frac{1}{2},3n+1;\frac{3}{2};1-\frac{R_{max}}{R_{min}}\right)
\end{equation}

\subsection{Hyperbolic Tube} \label{}

For a tube of hyperbolic profile, similar to the profile in Figure \ref{Parabola}, the radius is
given by

\begin{equation}
r(x)=\sqrt{a+bx^{2}} \hspace{1cm} -L/2 \le x \le L/2 \hspace{1cm} a,b>0
\end{equation}
where
\begin{equation}
 a=R_{min}^{2} \hspace{1cm} {\rm and} \hspace{1cm} b=\left(\frac{2}{L}\right)^{2}(R_{max}^{2}-R_{min}^{2})
\end{equation}

Therefore, Equation \ref{PQEq} becomes

\begin{eqnarray}
  P&=&\frac{2CQ^{n}(3n+1)^{n}}{\pi^{n}n^{n}}\int_{-L/2}^{L/2}\frac{dx}{\left(a+bx^{2}\right)^{(3n+1)/2}} \\
  &=&\frac{2CQ^{n}(3n+1)^{n}}{\pi^{n}n^{n}}\left[x\left(\frac{\frac{bx^{2}}{a}+1}{a+bx^{2}}\right)^{\frac{3n+1}{2}}{}_{2}F_{1}\left(\frac{1}{2},\frac{3n+1}{2};\frac{3}{2};-\frac{bx^{2}}{a}\right)\right]_{-L/2}^{L/2}
\end{eqnarray}
where $_{2}F_{1}$ is the hypergeometric function. Therefore

\begin{equation}\label{fHyperbolic}
P=\frac{2LCQ^{n}(3n+1)^{n}}{\pi^{n}n^{n}R_{min}^{3n+1}}\,_{2}F_{1}\left(\frac{1}{2},\frac{3n+1}{2};\frac{3}{2};1-\frac{R_{max}^{2}}{R_{min}^{2}}\right)
\end{equation}

\subsection{Hyperbolic Cosine Tube} \label{}

For a tube of hyperbolic cosine profile, similar to the profile in Figure \ref{Parabola}, the
radius is given by

\begin{equation}
r(x)=a\cosh(bx) \hspace{1cm} -L/2 \le x \le L/2 \hspace{1cm} a>0
\end{equation}
where
\begin{equation}
 a=R_{min} \hspace{1cm} {\rm and} \hspace{1cm} b=\frac{2}{L}\arccosh\left(\frac{R_{max}}{R_{min}}\right)
\end{equation}

Hence, Equation \ref{PQEq} becomes

\begin{eqnarray}
  P&=&\frac{2CQ^{n}(3n+1)^{n}}{\pi^{n}n^{n}a^{3n+1}}\int_{-L/2}^{L/2}\frac{dx}{\cosh^{3n+1}(bx)} \\
  &=&\frac{2CQ^{n}(3n+1)^{n}}{3\pi^{n}n^{n+1}a^{3n+1}b}\left[\frac{\sinh(bx)}{\cosh^{3n}(bx)\sqrt{-\sinh^{2}(bx)}}\,_{2}F_{1}\left(\frac{1}{2},-\frac{3n}{2};\frac{2-3n}{2};\cosh^{2}(bx)\right)\right]_{-L/2}^{L/2}
\end{eqnarray}

On evaluating the last expression and taking its real part which is physically significant, the
following relation is obtained
\begin{equation}\label{fCoshine}
P=\frac{2LCQ^{n}(3n+1)^{n}}{3\pi^{n}n^{n+1}R_{min}R_{max}^{3n}\arccosh\left(\frac{R_{max}}{R_{min}}\right)}\mathrm{Im}\left(_{2}F_{1}\left(\frac{1}{2},-\frac{3n}{2};\frac{2-3n}{2};\frac{R_{max}^{2}}{R_{min}^{2}}\right)\right)
\end{equation}
where $\mathrm{Im}(_{2}F_{1})$ is the imaginary part of the hypergeometric function.

\subsection{Sinusoidal Tube} \label{}

\begin{figure}[!h]
\centering{}
\includegraphics
[scale=0.8] {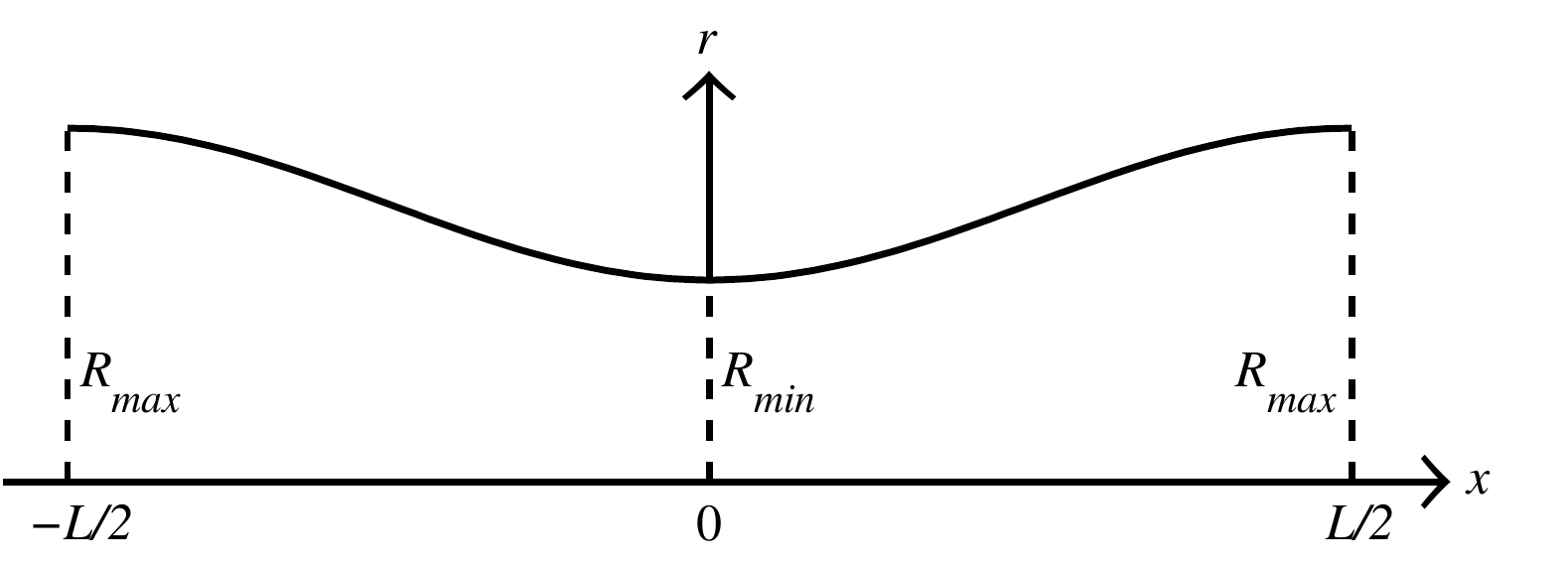} \caption{Schematic representation of the radius of a converging-diverging
capillary with a sinusoidal profile as a function of the distance along the tube axis.}
\label{Sinusoid}
\end{figure}

For a tube of sinusoidal profile, depicted in Figure \ref{Sinusoid}, where the tube length $L$
spans one complete wavelength, the radius is given by
\begin{equation}
r(x)=a-b\cos\left(kx\right) \hspace{1cm} -L/2 \le x \le L/2 \hspace{1cm} a>b>0
\end{equation}
where
\begin{equation}
a=\frac{R_{max}+R_{min}}{2} \hspace{1cm} b=\frac{R_{max}-R_{min}}{2} \hspace{1cm} \& \hspace{1cm}
k=\frac{2\pi}{L}
\end{equation}

Hence, Equation \ref{PQEq} becomes
\begin{eqnarray}
  P&=&\frac{2CQ^{n}(3n+1)^{n}}{\pi^{n}n^{n}}\int_{-L/2}^{L/2}\frac{dx}{\left[a-b\cos\left(kx\right)\right]^{3n+1}} \\
  &=&-\frac{2CQ^{n}(3n+1)^{n}}{3\pi^{n}n^{n+1}k}\left[\frac{1}{\sin(kx)\left[a-b\cos\left(kx\right)\right]^{3n}}\sqrt{\frac{-\sin^{2}(kx)}{a^{2}-b^{2}}}\,\mathfrak{A}\right]_{-L/2}^{L/2}
\end{eqnarray}
where $\mathfrak{A}$ is the Appell hypergeometric function given by

\begin{equation}
\mathfrak{A}=F_{1}\left(-3n;\frac{1}{2},\frac{1}{2};1-3n;\frac{a-b\cos(kx)}{a+b},\frac{a-b\cos(kx)}{a-b}\right)
\end{equation}

On taking the limit as $x\rightarrow-\frac{L}{2}^{+}$ and $x\rightarrow\frac{L}{2}^{-}$ and
considering the real part, the following relation is obtained
\begin{equation}\label{fSinusoid}
P=\frac{2LCQ^{n}(3n+1)^{n}}{3\pi^{n+1}n^{n+1}R_{max}^{3n}\sqrt{R_{max}R_{min}}}\mathrm{Im}\left(F_{1}\left(-3n;\frac{1}{2},\frac{1}{2};1-3n;1,\frac{R_{max}}{R_{min}}\right)\right)
\end{equation}
where $\mathrm{Im}(F_{1})$ is the imaginary part of the Appell hypergeometric function.

\vspace{0.5cm}

It is noteworthy that all these relations (i.e. Equations \ref{fConic}, \ref{fParabolic},
\ref{fHyperbolic}, \ref{fCoshine} and \ref{fSinusoid}), are dimensionally consistent. Moreover,
they have been thoroughly tested and validated by numerical integration.

\newpage
\section{Conclusions} \label{Conclusions}

In this article a mathematical method for obtaining analytical relations between the flow rate and
pressure drop in axisymmetric corrugated tubes is presented and applied to the flow of power-law
fluids. The method is illustrated by five examples of circular capillaries with
converging-diverging shape. This method can be used to derive similar relations for other types of
fluid and other types of tube geometry. It can also be used as a base for numerical integration
where analytical relations are difficult to obtain due to mathematical or practical reasons.

\vspace{1cm}

\newpage
\phantomsection \addcontentsline{toc}{section}{Nomenclature} %
{\noindent \LARGE \bf Nomenclature} \vspace{0.5cm}

\begin{supertabular}{ll}
$\sR$                 & strain rate (s$^{-1}$) \\
$\mu$                 & fluid viscosity (Pa.s) \\
$\sS$                 & stress (Pa) \\
\\
$C$                   & consistency factor (Pa.s$^n$) \\
$F_1$                 & Appell hypergeometric function \\
$_2F_1$               & hypergeometric function \\
$L$                   & tube length (m) \\
$n$                   & flow behavior index \\
$P$                   & pressure drop (Pa) \\
$Q$                   & volumetric flow rate (m$^{3}$.s$^{-1}$) \\
$r$                   & tube radius (m) \\
$R_{max}$             & maximum radius of corrugated tube (m) \\
$R_{min}$             & minimum radius of corrugated tube (m) \\
$x$                   & axial coordinate (m) \\
\end{supertabular}

\newpage
\phantomsection \addcontentsline{toc}{section}{References} %
\bibliographystyle{unsrt}

\end{document}